\documentclass[12pt]{article}
\usepackage{graphicx}

\newcommand\pubnumber{DPF2013-116}
\newcommand\pubdate{\today}

\def\berkeley{Lawrence Berkeley National Laboratory\\
1 Cyclotron Rd., 94720 Berkeley (CA), USA}
\def\support{\footnote{Work partially supported by the Director, Office of Science, Office of Basic Energy Sciences, of the U.S. Department of Energy under Contract No. DE-AC02-05CH11231.}}

\def\Title#1{\begin{center} {\Large #1 } \end{center}}
\def\Author#1{\begin{center}{ \sc #1} \end{center}}
\def\Address#1{\begin{center}{ \it #1} \end{center}}

\newcommand\pubblock{\rightline{\begin{tabular}{l} \pubnumber\\
         \pubdate  \end{tabular}}}
\newenvironment{Abstract}{\begin{quotation}  }{\end{quotation}}
\newenvironment{Presented}{\begin{quotation} \begin{center} 
             PRESENTED AT\end{center}\bigskip 
      \begin{center}\begin{large}}{\end{large}\end{center} \end{quotation}}
\def\Acknowledgments{\bigskip  \bigskip \begin{center} \begin{large}
             \bf ACKNOWLEDGMENTS \end{large}\end{center}}

\def\beq{\begin{equation}}
\def\eeq#1{\label{#1}\end{equation}}
\def\eeqn{\end{equation}}

\def\beqa{\begin{eqnarray}}
\def\eeqa#1{\label{#1}\end{eqnarray}}
\def\eeqan{\end{eqnarray}}

\let\bar=\overbar

\def\Dslash{\not{\hbox{\kern-4pt $D$}}}
\def\dslash{\not{\hbox{\kern-2pt $\del$}}}

\def\msb{{\bar{\ssstyle M \kern -1pt S}}}

\begin{document}
\begin{titlepage}
\pubblock

\vfill
\Title{Silicon strip prototypes for the Phase-II upgrade of the ATLAS tracker for the HL-LHC}
\vfill
\Author{ Sergio D\'iez\support}
\Address{\berkeley}
\vfill
\begin{Abstract}
This paper describes the integration structures for the silicon strips tracker of the ATLAS detector proposed  for the Phase-II upgrade of the Large Hadron Collider (LHC), also referred to as High Luminosity LHC (HL-LHC). In this proposed detector Silicon strip sensors are arranged in highly modular structures, called `staves' and `petals'. This paper presents performance results from the latest prototype stave built at Berkeley. This new, double-sided prototype is composed of a specialized core structure, in which a shield-less bus tape is embedded in between carbon fiber lay-ups. A detailed description of the prototype and its electrical performance is discussed in detail in this paper.
\end{Abstract}
\vfill
\begin{Presented}
DPF 2013\\
The Meeting of the American Physical Society\\
Division of Particles and Fields\\
Santa Cruz, California, August 13--17, 2013\\
\end{Presented}
\vfill
\end{titlepage}
\def\thefootnote{\fnsymbol{footnote}}
\setcounter{footnote}{0}

\section{Introduction}

At the next-generation tracking detector proposed for the High Luminosity LHC (HL-LHC), the so-called ATLAS Phase-II Upgrade, the particle densities and radiation levels will be higher by as much as a factor of ten. The new detectors must be faster, more highly segmented, cover more area, be more resistant to radiation, and require much greater power delivery to the front-end systems. At the same time, they cannot introduce excess material which could undermine performance. For those reasons, the inner tracker of the ATLAS detector must be redesigned and rebuilt completely \cite{Gianotti2005}. The design of the inner tracker for the Phase-II ATLAS Upgrade of HL-LHC has already been defined \cite{LoI}. It consists of several layers of silicon particle detectors. The innermost layers will be composed of silicon pixel sensors, and the outer layers will be composed of silicon "short" ($\sim 2.5\; cm$) and "long" ($\sim 5\; cm$) strip sensors. In response to the needs of the strip region for the upgraded tracker, highly modular structures are being studied and developed, called `staves' for the central region (barrel) and `petals' for the forward regions (end-caps). These structures integrate large numbers of sensors and readout electronics, with precision light weight mechanical elements and cooling structures. This papers focuses on the development of one of the latest barrel stave prototypes.

\begin{figure}[htb]
\centering
\includegraphics[width=.9\textwidth]{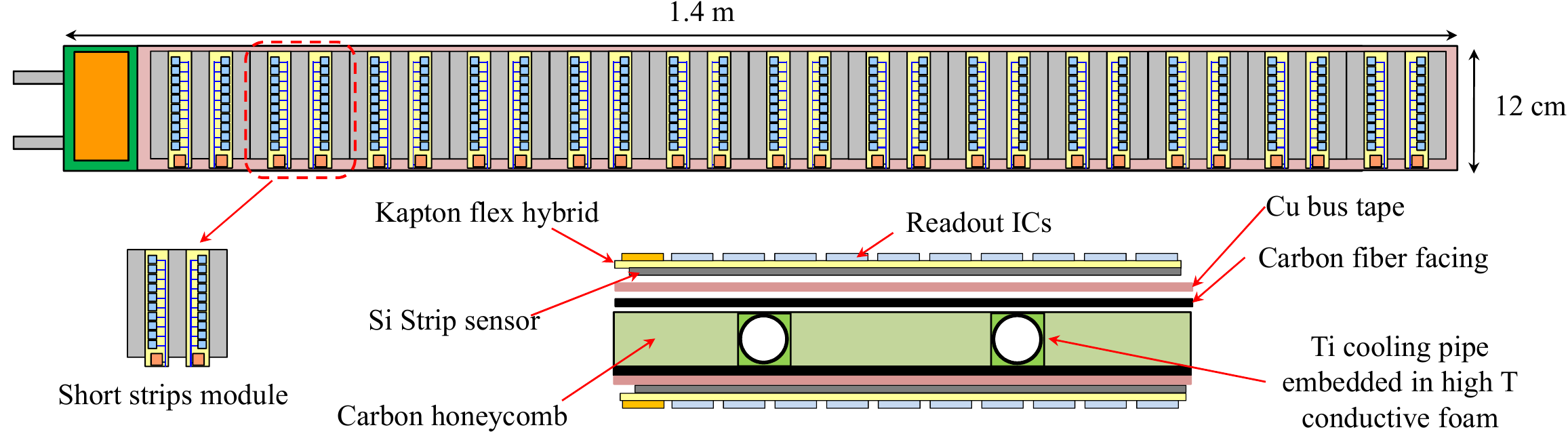}
\caption{Sketch of the baseline layout of a barrel short strips stave and its cross-section}
\label{figure:stavesketch}
\end{figure}

The baseline design of a stave consists of single-sided silicon strip sensors, glued onto a low mass carbon-based core and facings with embedded titanium cooling pipes. Readout, control, and power electronics are hosted in kapton flex hybrids, glued directly onto the silicon sensors. A stave hosts $13$ identical sensor modules per side, $26$ in total, having both axial and stereo strips providing 3D hit reconstruction. Figure~\ref{figure:stavesketch} shows a sketch of a short strip stave and its cross-section. Small versions of short strip staves with $4$ strip modules instead of $13$ (`stavelets') have already been developed by the strips collaboration \cite{Phillips2012}. This paper describes in detail the so-called shield-less stavelet, one of the latest stave prototypes recently assembled, with some unique features: it is the first double-sided stave prototype, with $4$ strip modules per side, and it includes a novel, low-mass bus tape and core structure.

\section{Description and manufacture}

\begin{figure}[htb]
\centering
\includegraphics[width=.45\textwidth]{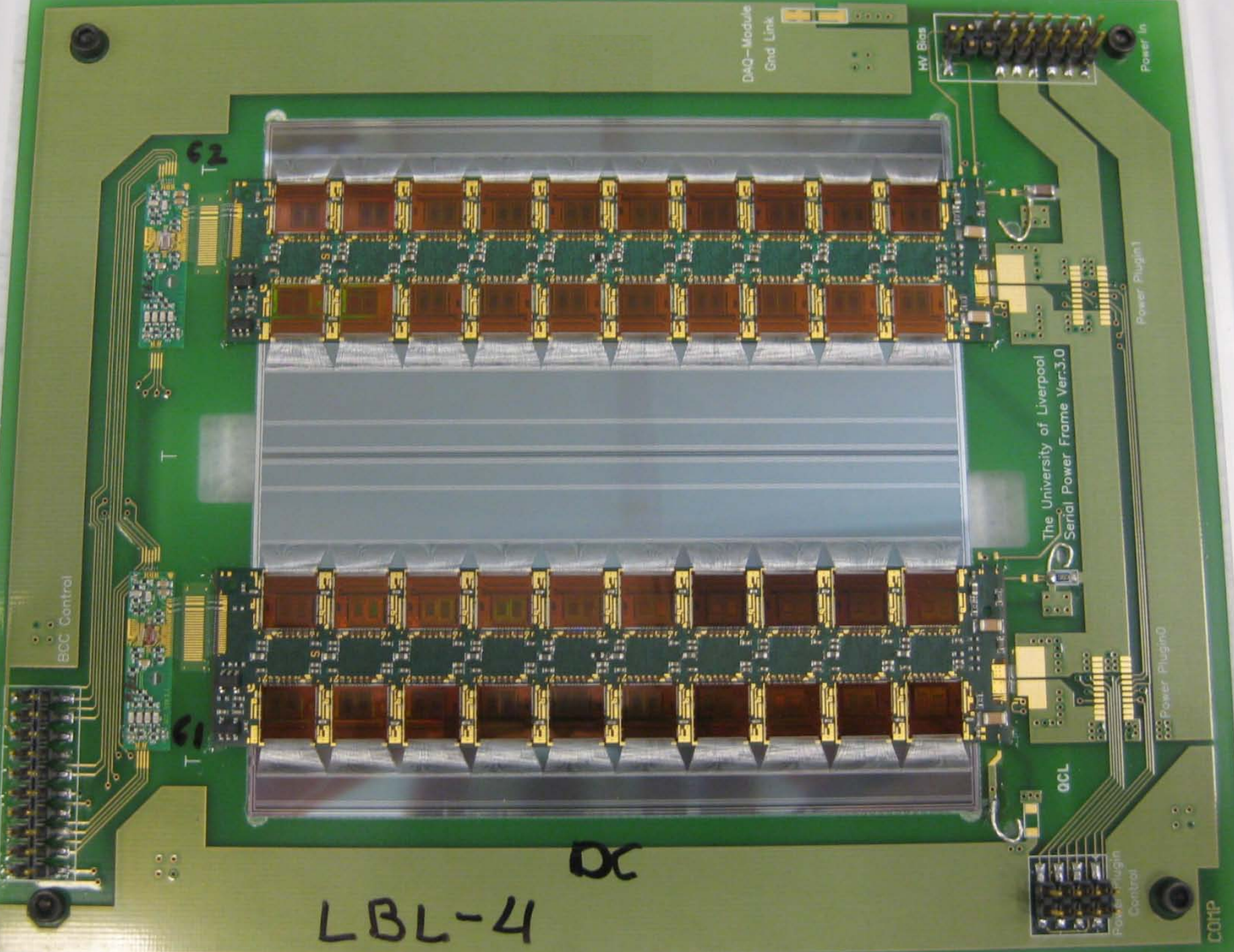}
\caption{Short strips module fully assembled and ready for test.}
\label{figure:module}
\end{figure}

The shield-less stavelet is the first double-sided stave prototype, with $4$ silicon microstrip modules per side, $8$ in total. One side has implemented the serial powering distribution, while the other side has the DC-DC powering distribution. Each module consists of a silicon microstrip sensor with integrated readout and control electronics. The silicon microstrip sensors are fabricated in n-in-p float zone (FZ) technology, consisting of a segmented $n^+$ implant on a p-bulk technology. The total size of a sensor is equal to $9.75\times9.75$ cm${}^2$, segmented in $4$ columns of $1280$ strips. Each strip is $2.39$ cm long, and the pitch between strips is equal to $74.5$ $\mu$m \cite{Unno2011}. Each short strip module includes two flex hybrids, which host the readout electronics and part of the power circuitry. The readout ASIC used in this prototype is the ABCN-25 binary readout ASIC, fabricated in $0.25$ $\mu$m CMOS technology \cite{Dabrowski2009}. Each ASIC reads out $128$ channels, and each channel is connected to its correspondent strip via wire bonds. Figure~\ref{figure:module} shows a silicon short strip barrel module fully assembled. Up to $9$ different institutes within the collaboration have assembled and tested more than $70$ barrel modules with high yield. The modules used in the shield-less stavelet were assembled at Berkeley Lab and University of California Santa Cruz. Further details about module construction and testing can be found in \cite{Allport2011, Diez2013}. 

The mechanical structure (`core') of the shield-less stavelet has several unique features; it includes a novel bus tape and core structure in which the thick ($\sim 50\; \mu$m) aluminum shield layer has been removed, as opposed to previous prototypes that followed the baseline layout more closely \cite{Phillips2012}. Once this Al layer is mostly removed, deformations during the tape manufacture and co-curing, observed in the baseline stavelet prototypes, become marginal. In addition, instead of co-curing the bus tape on top of the carbon fiber (CF) facings of the core as in the baseline layout, the shield-less tapes are co-cured in between two CF layers. The last CF skin, with its size reduced along the edges, acts as an effective shielding layer. Finally, the removal of the Al screen has additional advantages: the removal of a thick aluminum shielding layer implies a $\sim 50\%$ material reduction in the tapes, leading to an overall $10\%$ reduction of the stave radiation length; in addition, the simplified layout and bus tape manufacture leads to a cost reduction of the tapes. The Al shield layer was kept at the location of one of the modules per side, in order to have a straightforward comparison between shielded and shield-less modules. The remaining core structure consists of a `honeycomb' CF body with an embedded, U-shaped stainless steel cooling pipe, wrapped in thermally conductive carbon foam. The outer diameter of the cooling pipe is equal to $3.3$ mm, and the wall thickness is $220\;\mu$m. The core structure is sealed with the co-cured CF facings and tapes on both sides. The dimensions of the stavelet are $600$ mm long, $150$ mm wide, and $5$ mm thick. 

Two powering distributions are still under discussion for the staves: serial powering and DC-DC powering. The shield-less stavelet has both schemes implemented, one on each side. In serial powering, a constant current source provides power to all the modules of a stave side with a shunt regulator circuit and a single power line. The current is constant along the chain, and is equal to the current required by an individual module (`chain of modules' implementation). There are as many voltage steps as elements in the chain, and protection circuitry is required to prevent open circuits in the serial power line. For that purpose, a dedicated circuit and board called power protection board (PPB) was developed \cite{Lynn2011}. Another approach is used in DC-DC powering, in which a constant voltage source provides power to the modules in a stave side after a voltage conversion step is performed by buck DC-DC converters. It also requires a single power line. Current increases along the stave, and the total current at the end of the stave is equal to the number of modules in the stave, multiplied by the current required per hybrid, and divided by the voltage conversion ratio. Low-noise control and protection circuitry is also required, and for that dedicated ASICs have been developed and are included in the DC-DC converter board \cite{Faccio2010}. Strong efforts have been dedicated by the strips community to the study and development of both powering architectures, each with advantages and drawbacks \cite{Diez2012pow}. The shield-less stavelet provides a new and very powerful test bench for these studies, since it allows the simultaneous test of both powering distributions within the same system and under the same test conditions.

\begin{figure}[ht]
\centering
\includegraphics[width=.7\textwidth]{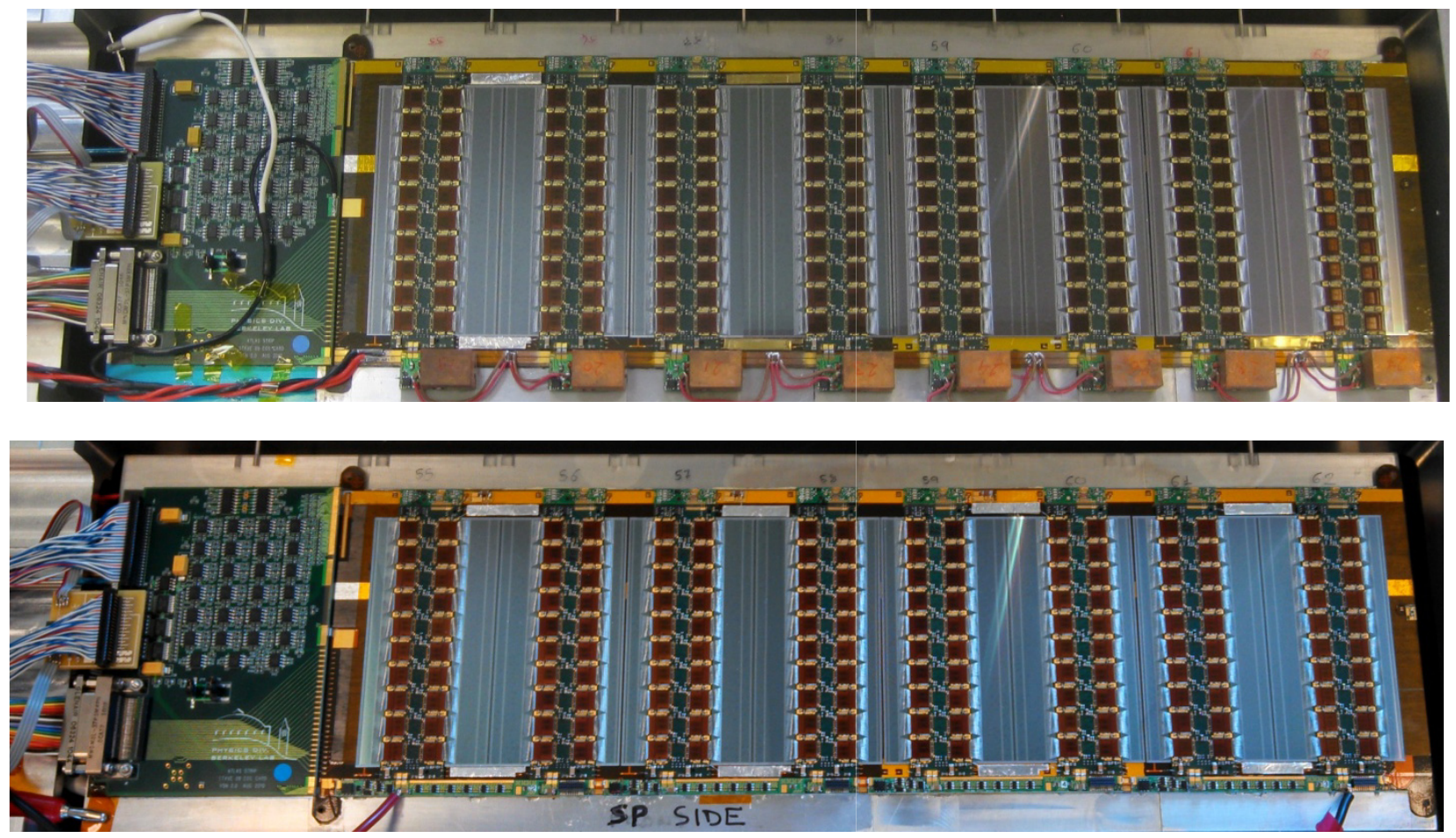}
\caption{Both sides of the shield-less stavelet. (Top) DC-DC power side. (Bottom) Serial power side.}
\label{figure:fullstavelet}
\end{figure}

The modules are directly glued onto the stavelet core with thermally conductive SE4445 epoxy from \textit{Dow Corning Inc.} \cite{Dowcorning}. The backplane contact of the silicon sensors with the HV pads of the tape is made with electrically conductive TRA-DUCT 2902 epoxy from \textit{Henkel Inc.} \cite{Henkel}. A glue mask determines the glue pattern on the sensor backplane, designed to maximize the SE4445 glue spread and hence the thermal conduction between the stavelet core and the strips modules. Precision mechanical tools were designed in order to glue the modules onto the stavelet core with the required precision. These tools allow a uniform, $\sim 175\;\mu m$ glue thickness, and a module position accuracy within $\sim150\;\mu m$ for the $X$ and $Y$ dimensions, well above the minimum precision required for this prototype ($500\;\mu m$). Figure~\ref{figure:fullstavelet} shows both sides of the stavelet fully assembled and ready for test.

\section{Prototype performance}
\label{sec:test}

The baseline DAQ system designed for that purpose is the High Speed Input/Output (HSIO) board \cite{NelsonHSIO}. It consists of a generic DAQ board with a single Virtex-4 Field Programmable Gate Array (FPGA) that allows data processing and connection to a controller PC. Up to $64$ simultaneous data streams are supported with this system. The setup also uses an upgraded version of the \textit{sctdaq} software, used in the past for the test of the current Semiconductor Tracker (SCT) modules \cite{Sctdaq2002}. During operation, the stavelet runs cooled down at $6{}^oC$ with a water chiller, and is kept inside a test box flooded with nitrogen, in order to shield the sensors from light and to prevent moisture to appear on the sensors. During normal operation, the DC-DC side runs at $V=10.5$ V and $I=10.3$ A, while the serial power side runs at $V=11.3$ V and $I=9.5$ A. 

\begin{figure}[ht]
\centering
\includegraphics[width=.99\textwidth]{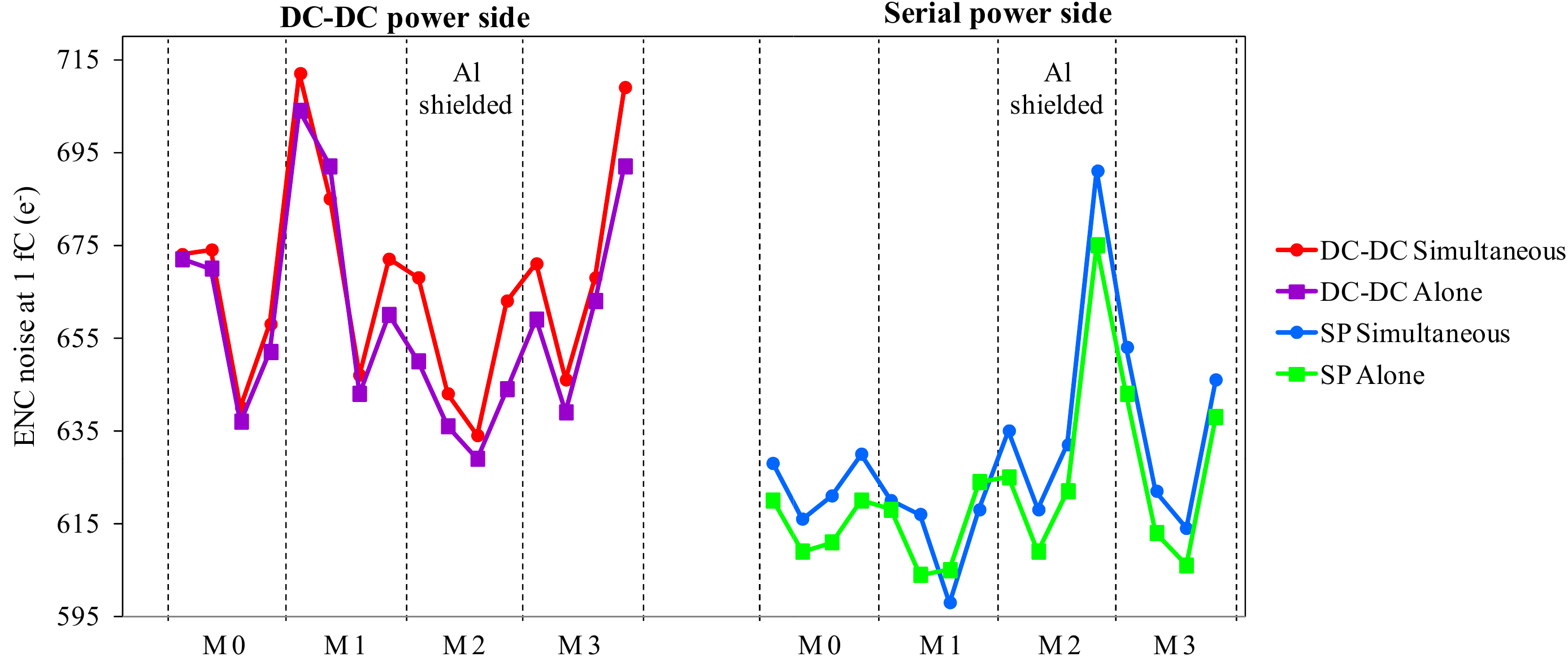}
\caption{ENC noise at $1\;fC$ injected charge versus the different modules of the shield-less stavelet. Both sides of the stavelet are represented. Each point represents the average noise of a chip column ($4$ columns per module). The location of the Al shielded modules is also indicated in the figure.}
\label{figure:ENC}
\end{figure}

The test setup described above allows the standard threshold scans for binary readout systems. One of the typical figures of merit of these tests is the input noise (or Equivalent Noise Charge, ENC, given in electrons) for each readout channel. It is obtained from the occupancy curves (`s-curves') of each channel at a particular injected charge. Figure ~\ref{figure:ENC} shows the ENC noise results at a $1\;fC$ injected charge for both sides of the shield-less stavelet (serial power and DC-DC power). Two types of results are shown: those obtained when both sides are being read out simultaneously and synchronously, and those obtained when each side is operated independently, with the other side turned off. The location of the Al shielded modules, one per side, is also indicated in the figure. 
Both sides show ENC noise values in the range of $595$ to $715\;e^-$. These values are within the expected range and are comparable with previous stavelet prototypes \cite{Phillips2012}. The Al shielded module does not exhibit significantly better ENC performances than the shield-less modules in neither of both stavelet sides. This is also true when both sides operate simultaneously. In addition, a consistently lower ENC noise can be observed in the serial power side of the stavelet as opposed to the DC-DC side, between $25$ and $30\; e^-$ on average (with the exception of one chip column of the shielded module). This noise difference is linked to the different power distributions of each side. Finally, results show a minor increase of ENC noise (around $7\;e^-$ on average) when both sides run simultaneously and synchronously. This slight noise increase can be fully attributable to the increase in temperature that each side overcomes when operating both sides at the same time (approximately $7{}^oC$), and not to noise interferences between both sides.

\begin{figure}[ht]
\centering
\includegraphics[width=.99\textwidth]{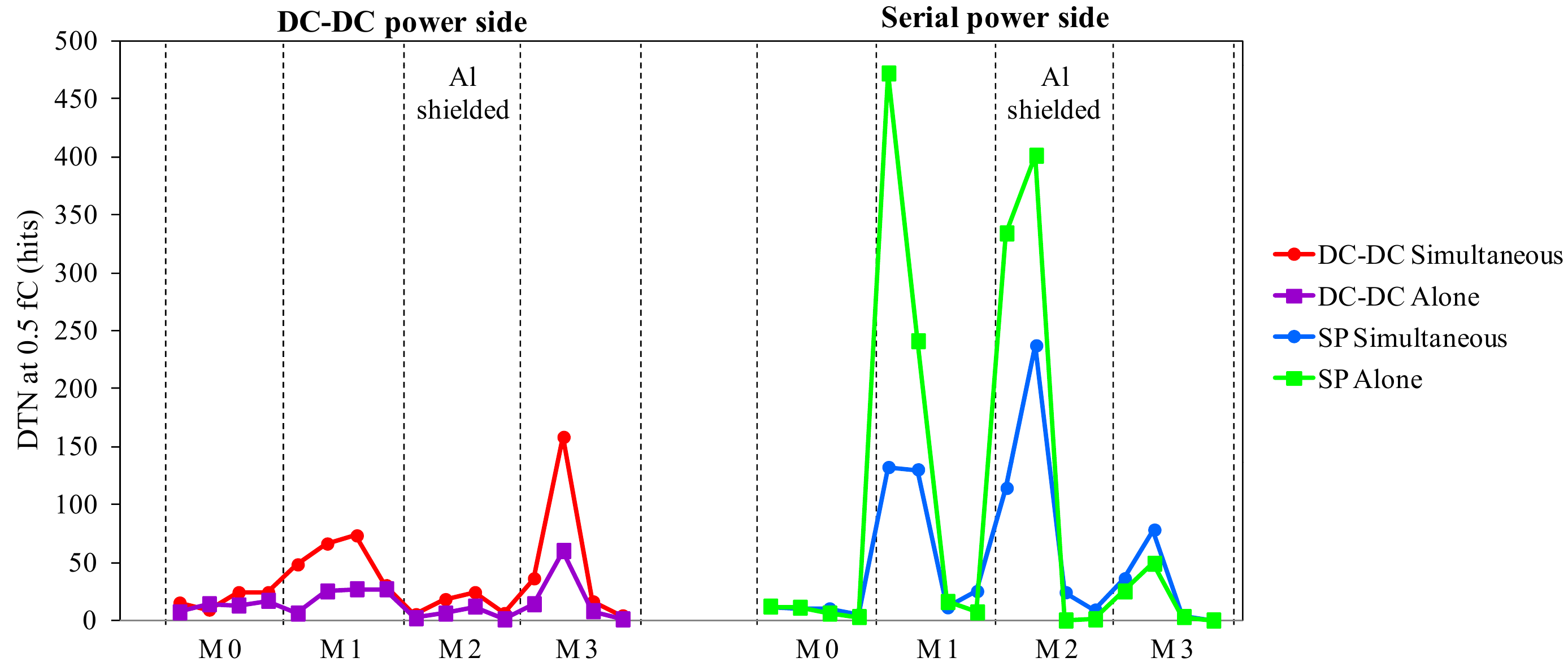}
\caption{Double trigger noise for an equivalent $V_T$ corresponding to a $0.5\;fC$ injected charge versus the different modules of the shield-less stavelet. Both sides of the stavelet are represented. Each point represents the total number of hits of a chip column ($4$ columns per module), for all the time interval.}
\label{figure:DTN}
\end{figure}

A well known issue for charge measurement systems is that readout trigger signals may produce noise interference during simultaneous charge integration. In order to probe that effect, the so-called double trigger noise (DTN) test was implemented on \textit{sctdaq}. It consists of the following: for a particular set of fixed thresholds, a first trigger signal is sent to the readout electronics. Then, after a controlled number of clock cycles, equal or close to the length of the cell pipeline of the readout ASICs, a second trigger is sent. The readout data obtained with the second trigger is looking at the charge integration occurred exactly when the first trigger signal was sent. This test is performed at three fixed threshold voltage values ($V_T$), corresponding to the $V_T$ that an equivalent injected charge equal to $0.5$, $0.75$, and $1\;fC$ would exhibit during a threshold scan. Since no charge is injected, ideally the number of hits read out at that particular time stamp should be zero. Figure~\ref{figure:DTN} shows the total number of hits produced per chip column after a double trigger noise test, for a time interval of $30$ clock cycles ($120$ to $150$ clock cycles), and at a $V_T$ corresponding to a $0.5\;fC$ injected charge.
$100$ measurements per channel are sent at each step. Results are again shown for both sides, running independently and simultaneously. Results show less than $100$ hits in most cases, with the exception of a single chip column in the DC-DC side and two hybrids in the serial power side. Big variations are observed from test to test ($\pm100$ hits) in the noisiest columns of the serial power side, that is, in the hybrids of module 1 and 2. The worst case observed is represented in this plot. The DC-DC side, however, exhibits lower DTN values than the serial power side, and more consistent results from test to test. These effects are related to high-frequency common-mode noise being developed along the stavelet LVDS data lines. This is of critical importance for the serial power distribution, in which each module stands at a different potential than the next. Results obtained with the independent readout of both sides exhibit very little differences with respect to the results obtained with both sides running at the same time and synchronously: this proves that the digital activity of one side does not interfere with the other. Again, very little differences are observed between shielded and shield-less modules. Although not shown here, DTN results obtained at $V_T$ values corresponding to $0.75$ and $1\;fC$ injected charges exhibit no hits on either case all along the time interval measured.

\section{Conclusions}

The first shield-less, double-sided stavelet prototype for the strips region of the ATLAS upgraded tracker has been built and tested. The electrical performances are comparable to the ones obtained with previous single-sided stavelet prototypes. The prototype allows a straightforward comparison between both powering distributions, serial powering and DC-DC powering: slightly better ENC noise performances are observed in the serial powering side, while the DC-DC powering side exhibits higher DTN immunity than the serial power side. In addition, no differences are observed between shielded or shield-less modules. All these results prove the robustness of the stave design as a double-sided prototype, and demonstrate that the Al screen on the bus tape is not required. The removal of the Al layer allows a significant material reduction in the prototypes, a reduction in costs of the bus tapes manufacture, and a simplified build procedure of the stavelet mechanical cores.

\Acknowledgments
The authors would like to thank the undergraduate students A. Faroni and T. Txiao, from University of California Berkeley (USA), and N. Lehman and M. Defferrard, from University of Fribourg (Switzerland) for his dedication and efficient work at Berkeley Lab during the different stages of the project.

\end{document}